\documentclass[11pt]{article}
\usepackage{moriond,epsfig}

\def\Journal#1#2#3#4{{#1} {\bf #2}, #3 (#4)}

\def\PRL{\em Phys. Rev. Lett.}
\def\PRD{{\em Phys. Rev.} D}

\def\be{\begin{equation}}
\def\ee{\end{equation}}
\def\bea{\begin{eqnarray}}
\def\eea{\end{eqnarray}}

\begin{document}
\vspace*{4cm}
\title{Deeply Virtual Compton Scattering and Higgs Production Using the Pomeron in AdS}

\author{ Richard C. Brower\footnote{speaker}}

\address{Department of Physics, Boston University, Boston MA 02215}

\author{Miguel S. Costa}

\address{Centro de F\'isica do Porto, Universidade do Porto, 4169-007 Porto, Portugal}

\author{Marko Djuri\'c\footnotemark[\value{footnote}]}

\address{Centro de F\'isica do Porto, Universidade do Porto, 4169-007 Porto, Portugal}

\author{Chung-I Tan}

\address{Physics Department, Brown University, Providence RI 02912}

\maketitle\abstracts{
In the past decade overwhelming evidence has emerged for a conjectured duality between a wide class of gauge theories in $d$ dimensions and string theories on asymptotically  $AdS_{d+1}$ spaces. We apply this duality to scattering processes that occur via Pomeron exchange. First we develop the Pomeron in string theory, as done by Brower, Polchinski, Strassler and Tan,\cite{Brower:2006ea} showing that it naturally emerges as the Regge Trajectory of the $AdS$ graviton. Next we apply the $AdS$ Pomeron to the study of Deeply Virtual Compton Scattering (DVCS), and see that our model gives good results when compared to HERA data.\cite{Costa:2012fw} We then show how we can extend our results to double Pomeron exchange, and apply it to developing a formalism for the study of double diffractive Higgs production.\cite{Brower:2012mk}}

\paragraph{Pomeron-Graviton Duality:}\label{sec:pomeron}

In the Regge limit, $s\gg t$, it can be shown for a wide range of scattering processes that the amplitude is dominated by Pomeron exchange. Traditionally this has been modeled at weak coupling using perturbative QCD, but we will use here a formulation based on gauge/gravity duality, or the $AdS/CFT$ correspondence, of which one particular example is the duality between $\mathcal{N} = 4$ SYM and Type IIB string theory on $AdS_5 \times S^5$. This approach has the advantages of allowing us to study the strong coupling region, providing a unified soft and hard diffractive mechanism, and as we will see it also fits well the experimental data. 

In lowest order in weak 't Hooft coupling for QCD, a bare Pomeron was first identified by Low and Nussinov as a two gluon exchange corresponding to a Regge cut in the $J$-plane at $j_0 = 1$.   Going beyond the leading order, Balitsky, Fadin, Kuraev and Lipatov (BFKL) summed all the diagrams for two gluon exchange to first order in $\lambda = g^2 N_c$ and {\em all} orders in $(g^2 N_c \log s)^n$, thus giving rise to the so-called BFKL Pomeron. The position of this $J$-plane cut is at $j_0 = 1+ \log (2) g^2 N_c /\pi^2$, recovering the Low-Nussinov result in the $\lambda\rightarrow 0$ limit. In a holographic approach to diffractive scattering~\cite{Brower:2006ea,Brower:2007qh,Brower:2007xg,Cornalba:2006xm}, the weak coupling Pomeron is replaced by the ``Regge graviton'' in AdS space, as formulated by Brower, Polchinski, Strassler and Tan (BPST)~\cite{Brower:2006ea,Brower:2007xg} which has both hard components due to near conformality in the UV and soft Regge behavior in the IR.  Corrections to the  strong coupling lower the intercept from $j=2$ to
\be
j_0 = 2 - 2 /\sqrt{g^2N_c}    \; .
\label{eq:BPST-intercept}
\ee
\begin{figure}
\begin{center}
\hskip .25cm 
\includegraphics[height=0.325 \textwidth,width=.425\textwidth]{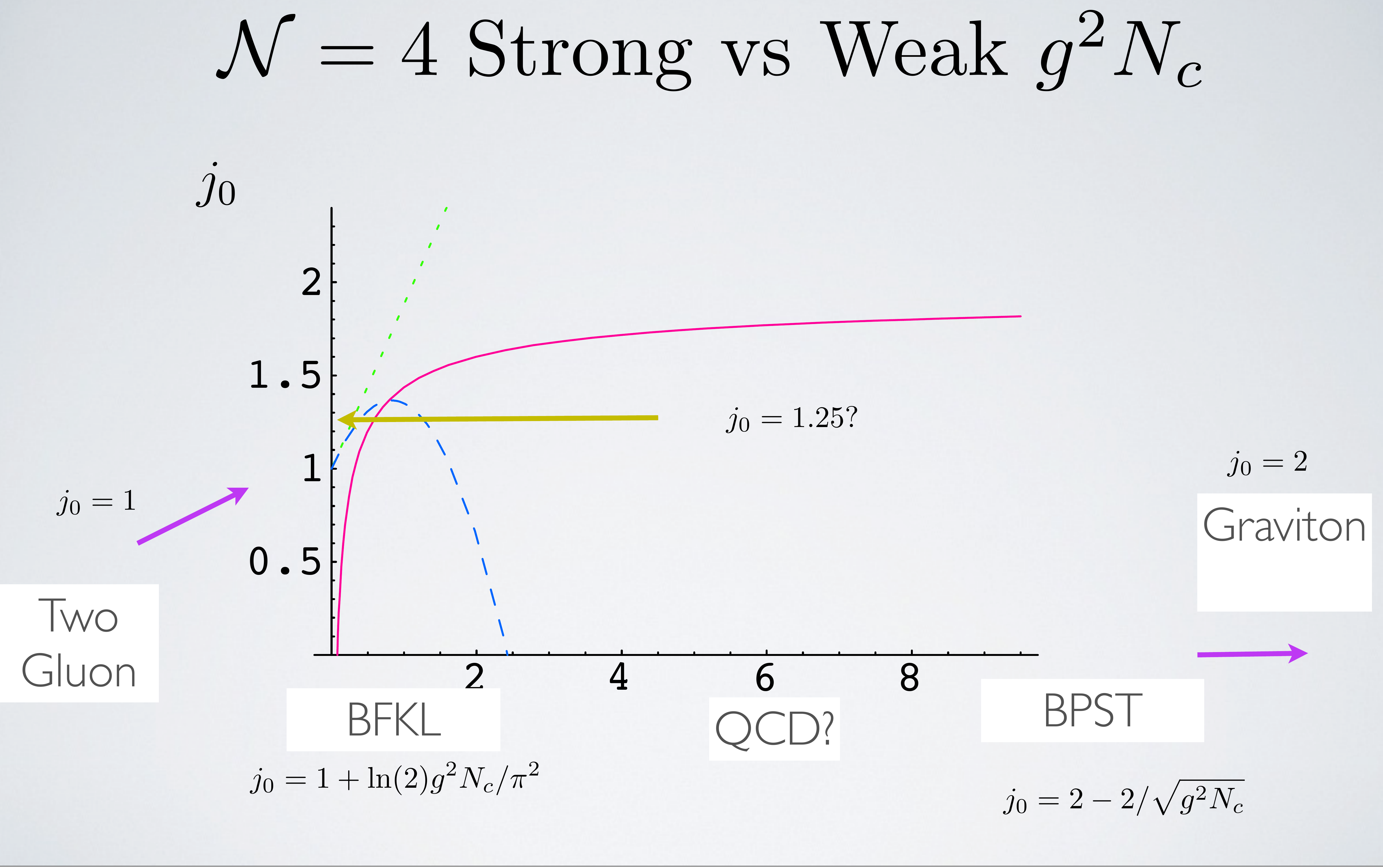}
\includegraphics[height=0.35 \textwidth,width=.45\textwidth]{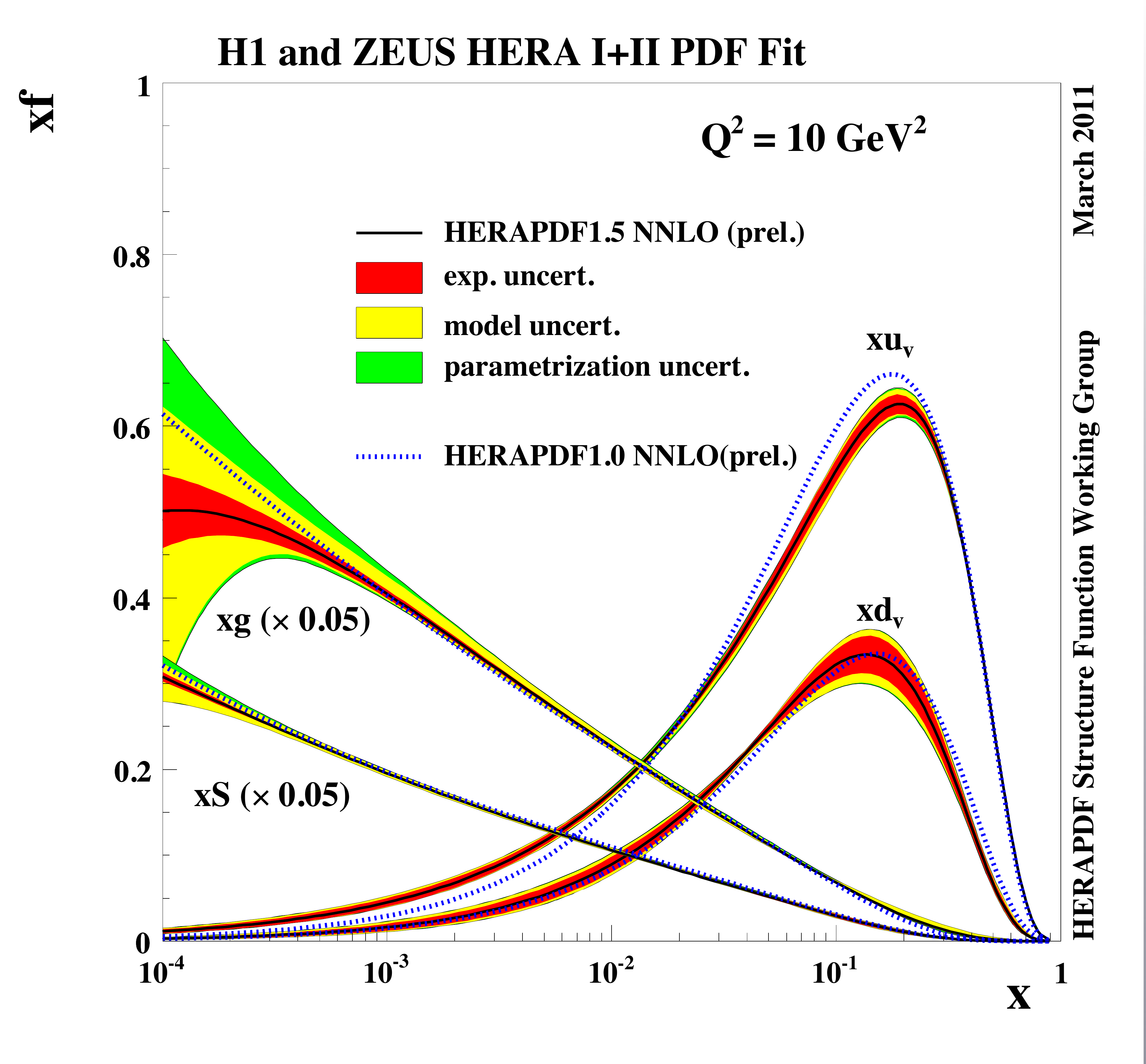}
\end{center}
\caption{ On the left,  intercept $j_0$ in ${\cal N}=4$ YM shown as a function of 't Hooft coupling $\lambda$ for the BPST Pomeron (solid red) and for BFKL (dotted and dashed to first and second order in $\lambda$ respectively). On the right, a typical partonic fit to HERA DIS data demonstrating the dominance for gluon dynamics at small $x$.  }
\label{fig:effective}
\end{figure}
 In Fig.~\ref{fig:effective},  we compare this  with the weak coupling BFKL intercept to second order.  A typical phenomenological estimates for this parameter  for
QCD is about $j_0 \simeq 1.25$,  which suggests that the physics of diffractive scattering is in the cross over region between
strong and weak coupling.  A corresponding treatment for Odderons has also been carried out~\cite{Brower:2008cy}. 
 We also show in Fig.~\ref{fig:effective} the dominance of gluons, in a conventional partonic approach, thus further justifying the large $N_c$ approximation, where  quark constituents are suppressed.

\paragraph{Holographic Treatment of Deeply Virtual Compton Scattering:}\label{sec:DVCS}

Previously, we have applied the $AdS/CFT$ correspondence to deep inelastic scattering\cite{Brower:2010wf,Cornalba:2010vk,Saturation} (see for example \cite{Brower:2011dx} for more references and the report from the previous Moriond conference). DVCS is the scattering between an off-shell photon and a proton, with the outgoing photon being on-shell. We  make use of the fact that the DVCS cross section and differential cross section can be related to the Pomeron exchange amplitude, $A(s,t)$ via 
\begin{equation}
\frac{d\sigma}{dt} (x,Q^2,t)= \frac{|A|^2}{16\pi s^2}   \,,\label{eq:diff_xsection}
\end{equation}
and
\begin{equation}
\sigma (x,Q^2)=\frac{1}{16\pi s^2}  \int dt \, |A|^2  \,. \label{eq:xsection}
\end{equation}
 In the holographic approach, the impact parameter space $(b_\perp, z)$ is 3 dimensional, where  $z \ge 0$ is the warped radial 5th dimension.  Conformal dilatations (
$\log z \rightarrow \log z + \mbox{const}$) take one from the UV boundary at $z = 0$ deep into the IR  $z = \mbox{large}$.    The near forward elastic amplitude  takes the 
eikonal form,
\begin{equation}
A(s,t)=2i s\int d^{2}b \; e^{i \vec q \cdot \vec b} \int dzdz'\; P_{13}(z)P_{24}(z')\big\{ 1- e^{i \chi(s,b,z,z')}\big\}  \; .\label{eq:A}
\end{equation}
where  $t=-q^2_{\perp}$ and the eikonal function,
$\chi$,
is related to a BPST Pomeron kernel in a transverse $AdS_3$ representation, $\mathcal{K}({s},b,z,z') $,  by 
\begin{equation}
\label{eq:chi_definition}
\chi( s,b,z,z')=\frac{g_{0}^{2}}{2{s}}(\frac{R^{2}}{zz'})^2\mathcal{K}({s},b,z,z').
\end{equation}
 An  important unifying features for the holographic map is factorization in the AdS space.  For hadron-hadron scattering, $P_{ij}(z)= \sqrt{-g(z)} (z/R)^2 \phi_i(z) \phi_j(z) $  involves a product of two external normalizable wave functions for the projectile and the target respectively.  For DVCS, states 1 and 3 are replaced by currents for an off-shell and on-shell photon respectively, and we can simply replace $P_{13}$ by product of the  appropriate unnormalized wave-functions. We can calculate these by evaluating the R-current - graviton Witten diagram in AdS, and we get 
\begin{equation}
P_{13}(z) = - C\, \frac{\pi^2}{6}\,z^3\, K_1(Q z). \label{eq:p13}
\end{equation}
Here $C$ is a normalization constant that can be calculated in the strict conformal limit. When expanded to first order in $\chi$, Eq. (\ref{eq:A}) provides the contribution from exchanging a single Pomeron. When $\chi$ is large equation (\ref{eq:A}) can be replaced by an AdS black disk model.\cite{Costa:2012fw}. In the conformal limit, a simple expression can be found. Confinement can next be introduced, eg., via a hardwall model $z < z_{cut-off}$. The effect of saturation can next be included  via the full transverse $AdS_3$ eikonal representation (\ref{eq:A}). 

\paragraph{Pomeron Kernel:}

The leading order BFKL Pomeron has  remarkable properties. It  enters into the first term in the large $N_c$ expansion with zero beta function.  Thus it is in effect the 
weak coupling cylinder graph for the  Pomeron for a large $N_c$  conformal theory, the same approximations used in the
AdS/CFT approach albeit at strong coupling. Remarkable BFKL integrability properties allows one to treat the BFKL  kernel 
as the solution to  an  $SL(2,\mathcal{C})$ conformal spin chain. Going to strong coupling, the  two gluon  exchange  evolves into a close string
of infinitely many tightly bound  gluons but the same underlying symmetry persists, referred to as  M\"obius invariance  in string theory or the
isometries of the transverse $AdS_3$ impact parameter geometry.  The position of the $j$-plane cut moves
from  $j_0 = 1+ \log (2) g^2 N_c/\pi^2$  up to $j_0 = 2- 2/\sqrt{g^2 N_c} $ and the kernel
obeys a Schr\"odinger equation on $AdS_3$ space for the Lorentz boost operators $M_{+-}$ ,
\begin{equation}
\left[ (-\partial_u^2 - te^{-2u})/2+\sqrt{\lambda}(j-j_0) \right]G_j(t,z,z')=\delta(u-u'),\label{adseq}
\end{equation}
with $z=e^{-u}.$   In the conformal limit,   $G_j(t,z,z')= \int dq\; q \; J_{\tilde \Delta(j)}(zq) J_{\tilde\Delta(j)}(qz')/(q^2-t)$, $\tilde\Delta(j)^2 = 2\lambda (j-j_0)$, and the Pomeron kernel is  obtained via an inverse Mellin transform. From here we can obtain $\chi$ using (\ref{eq:chi_definition}). The solution  for $\chi$ exhibits diffusion
\begin{equation}
\chi(\tau,L) =  (\cot(\frac{\pi\rho}{2}) + i) g_{0}^{2} e^{(1-\rho)\tau}\frac{L}{\sinh L} \frac{\exp(\frac{-L^{2}}{\rho\tau})}{(\rho \tau)^{3/2}},\label{strongkernel}
\end{equation}
in the "size" parameter $\log z$  for the exchanged closed string, analogous to the BFKL kernel  at weak coupling, with  diffusing taking place in  $\log(k_\perp)$,  the virtuality of the off shell gluon dipole. 
The diffusion constant  takes on  $\mathcal{D} = 2/\sqrt{g^2N_c}$ at strong coupling compared to $\mathcal{D}  = 7 \zeta(3) g^2 N_c/2 \pi^2 $ in weak coupling.
The close analogy between the weak and strong coupling Pomeron
suggests the development of a hybrid phenomenology leveraging plausible interpolations between the two extremes.

\paragraph{Fit to HERA Data}

\begin{figure}[htbp]
\label{fig:dvcs_fit}
\begin{center}
\includegraphics[scale=0.28]{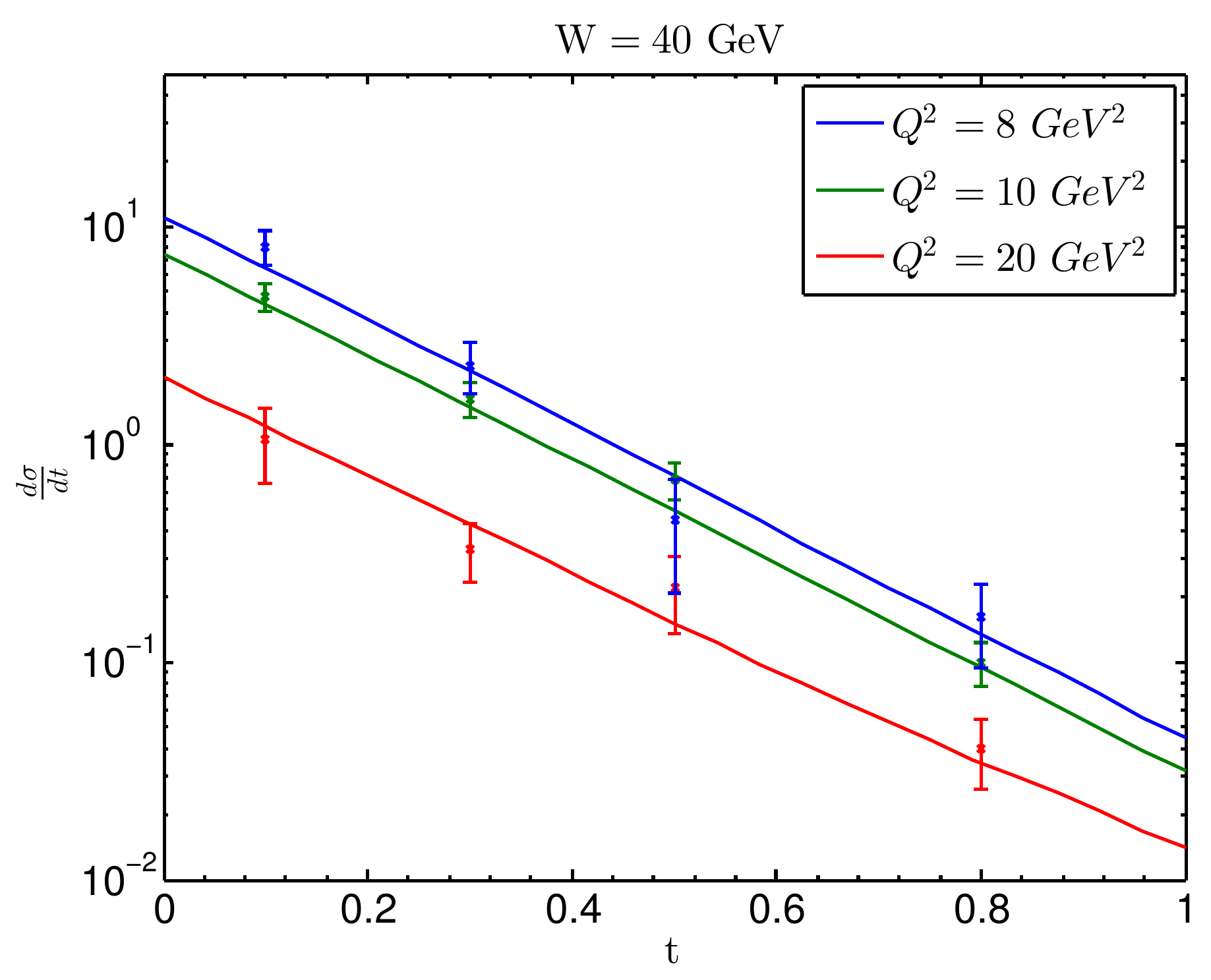}
\includegraphics[scale=0.28]{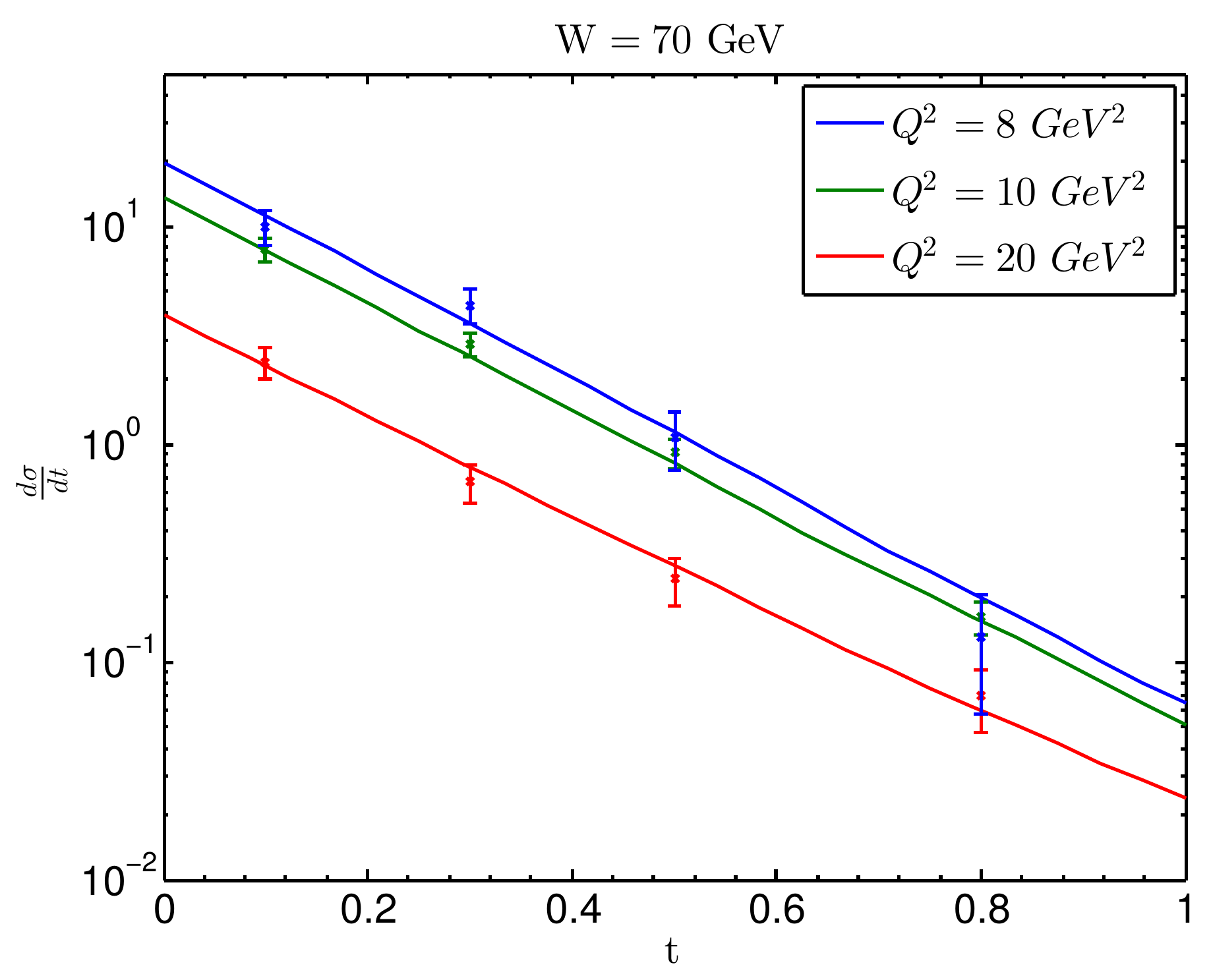}
\includegraphics[scale=0.28]{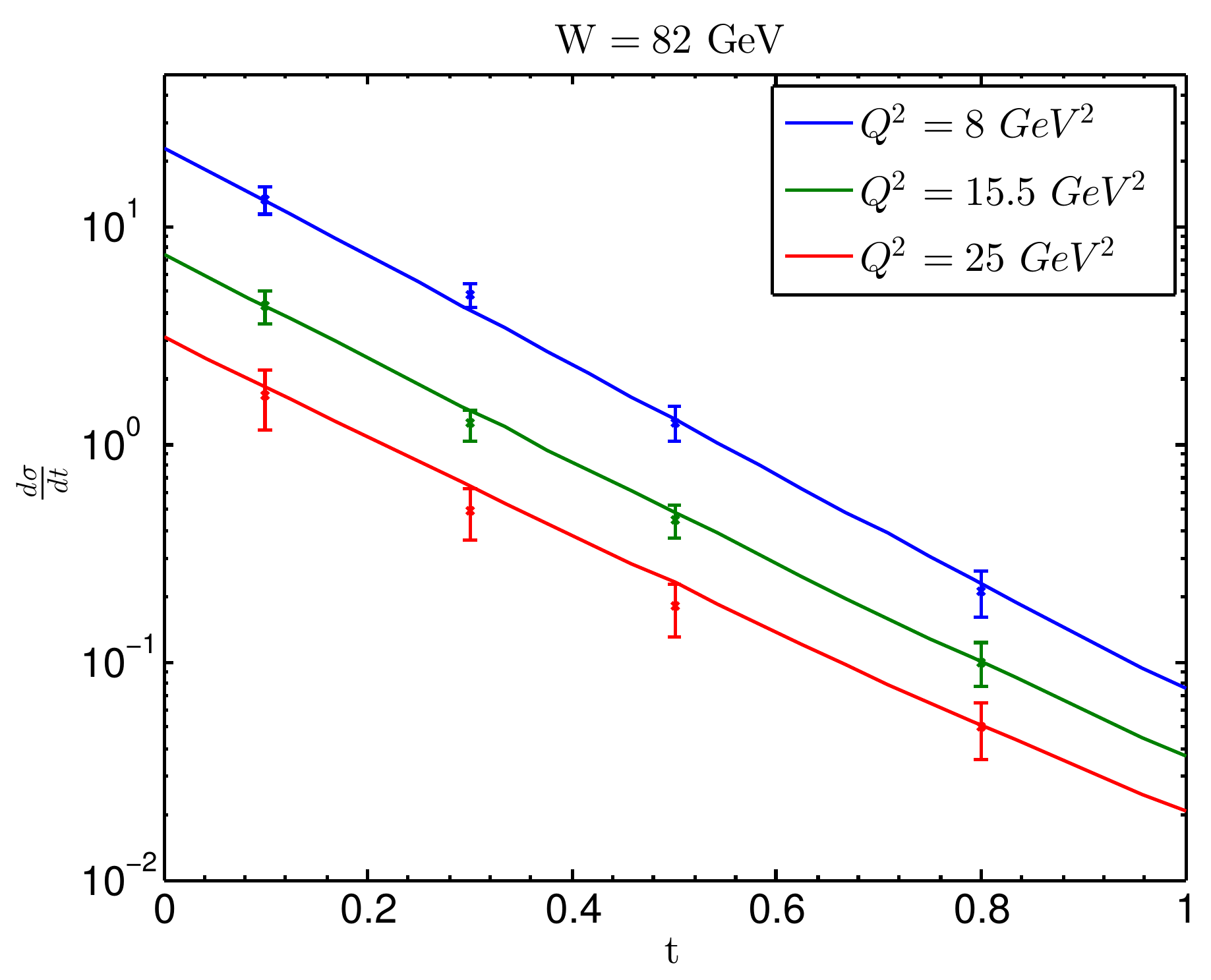}\\
\includegraphics[scale=0.28]{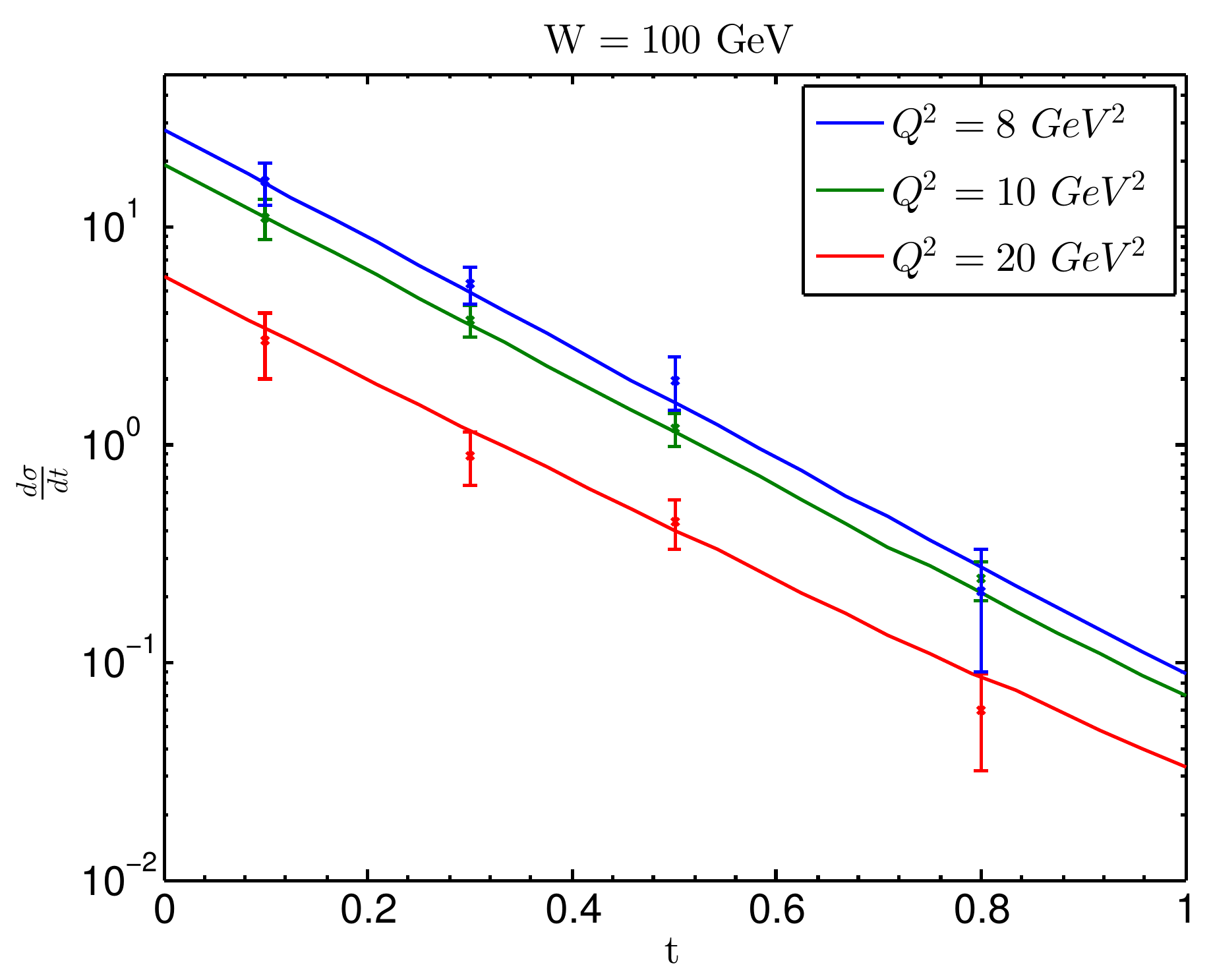}
\includegraphics[scale=0.28]{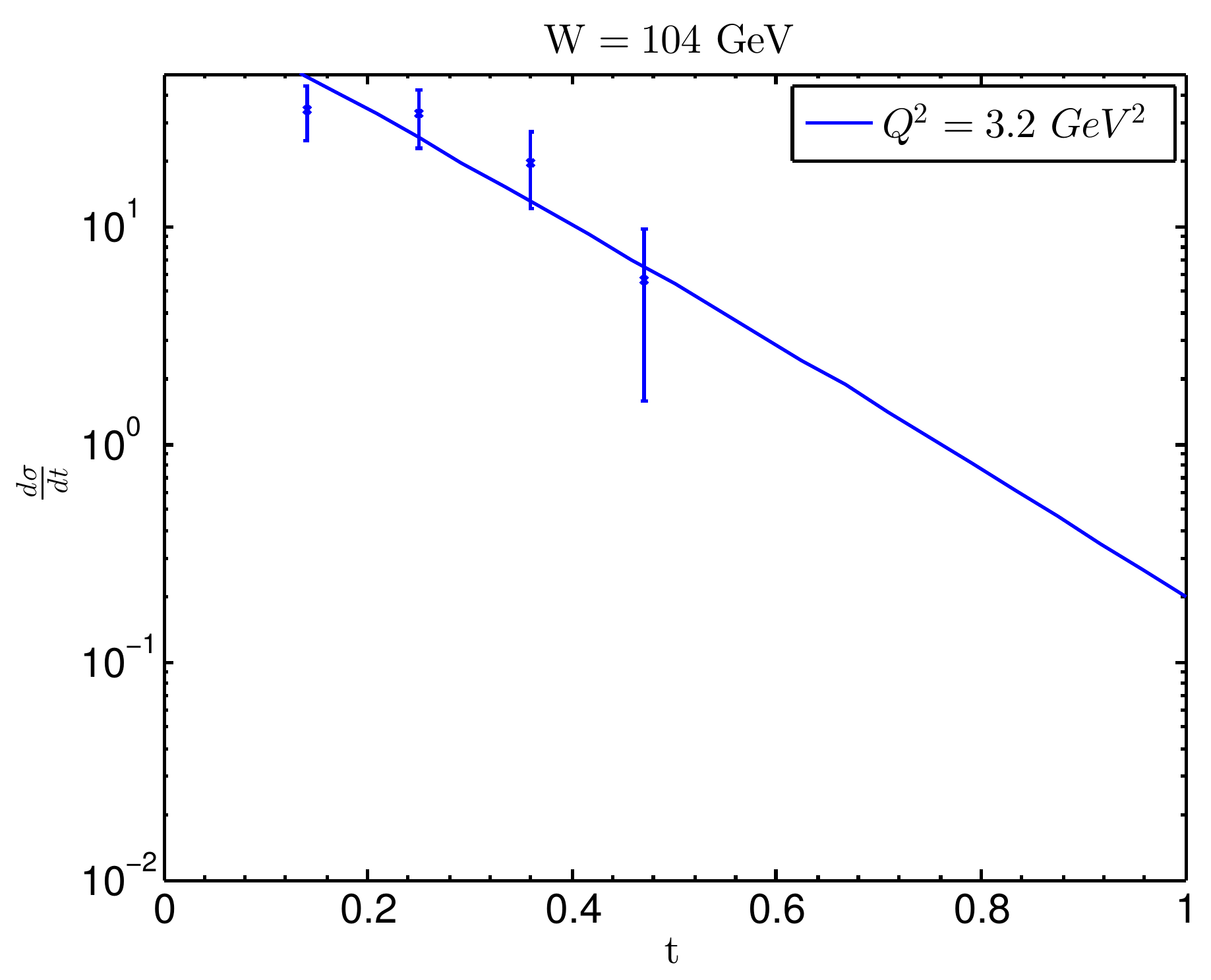}
\includegraphics[scale=0.28]{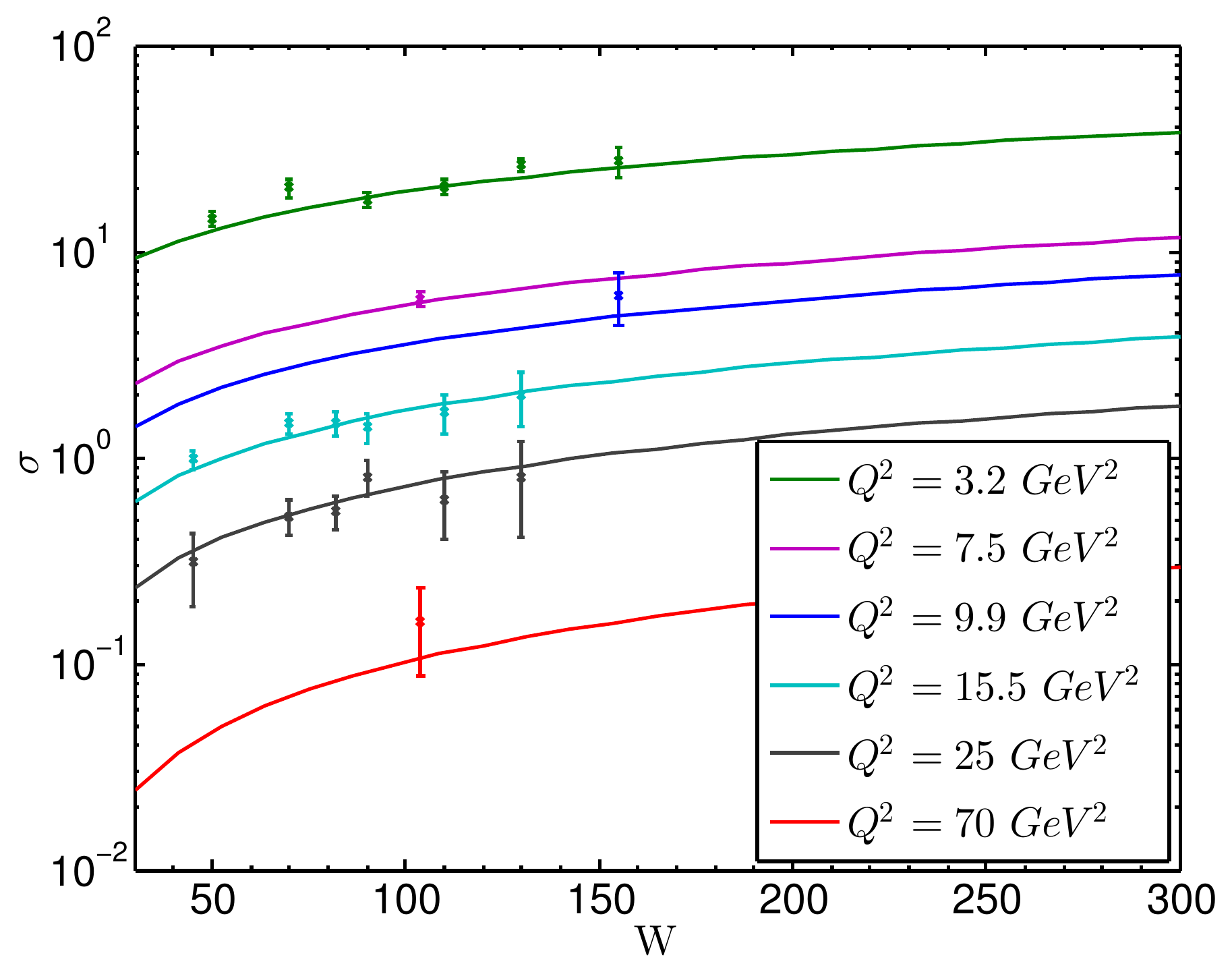}
\end{center}
\caption{The plots of the hard wall pomeron compared to HERA data. The first 5 correspond to the differential cross section, and the last one to the cross section where we omit some values of $Q^2$ to avoid cluttering the graph.}
\end{figure}

We now apply equations (\ref{eq:diff_xsection}) and (\ref{eq:xsection}) to compare our model to the measurements at HERA.\cite{Chekanov:2008vy,:2009vda} Related papers using AdS/CFT correspondence applied to DVCS include \cite{Gao:2009se,Marquet:2010sf,Nishio:2011xz}. We use equation (\ref{eq:p13}) for the photon wavefunctions and a delta function for the proton. Note that equation (\ref{strongkernel}) is for the conformal model, and the hard wall expression would include another term with the contribution due to the presence of the hard wall. See \cite{Costa:2012fw} for the explicit form. We obtain a good agreement with experiment, with $\chi^2$ varying from $0.51 - 1.33$ depending on the particular data and model we are considering. We find that confinement starts to play a role at small $|t|$, and the hardwall fits the data better in this region. Explicitly, the parameter values we get for the hard wall model are $g_0^2 = 2.46\, \pm  0.70\,, z_* = 3.35 \pm 0.41\ {\rm GeV}^{-1},\ \rho = 0.712 \pm 0.038\,,\ z_0 = 4.44 \pm 0.82\ {\rm GeV}^{-1}$ for the differential cross section, and $g_0^2 = 6.65 \pm 2.30\,, z_* = 4.86 \pm 2.87\ {\rm GeV}^{-1} ,\ \rho = 0.811 \pm 0.036\,,\  z_0 = 8.14 \pm 2.96\ {\rm GeV}^{-1},$ with $\chi^2_{d.o.f}=0.51$ and $1.03$ respectively. In figure \ref{fig:dvcs_fit} we present the plots corresponding to these parameters.

\paragraph{Double Diffractive Higgs Production}

We would now like to extend these methods to double diffractive Higgs production from forward proton-proton scattering, $p p \rightarrow p H p$.\cite{Brower:2012mk} The protons scatter through very
small angles with a large rapidity gaps separating the Higgs in the
central region. The Higgs subsequently decays into large transverse momentum
fragments. Although this represents a small fraction of the total
cross section, the exclusive channel should provide an exceptional
signal to background discrimination by constraining the Higgs mass to both the energy of
decay fragments and the energy lost to the forward protons~\cite{Kharzeev:2000jwa}. To extend our previous methods to this process, first notice that after expanding equation (\ref{eq:A}) to single pomeron exchange, we can schematically represent it as
\begin{equation}
A(s,t)  = \Phi_{13}*\widetilde {\cal K}_P * \Phi_{24} \; .
\label{eq:adsPomeronScheme}
\end{equation}

\begin{figure}[h]
\qquad
\includegraphics[height=0.15 \textwidth,width=0.4\textwidth]{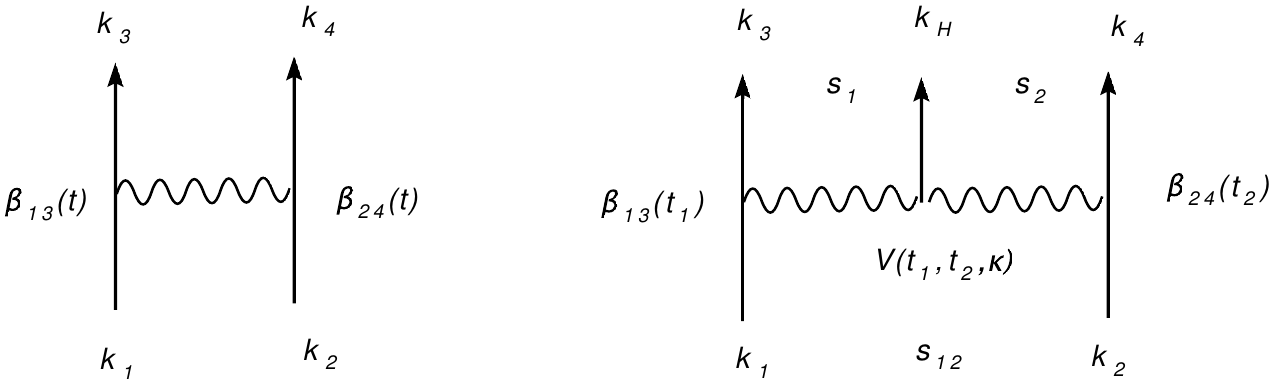}
\qquad
\qquad
\qquad
\includegraphics[angle = 90, height = 0.15\textwidth, width = 0.25\textwidth]{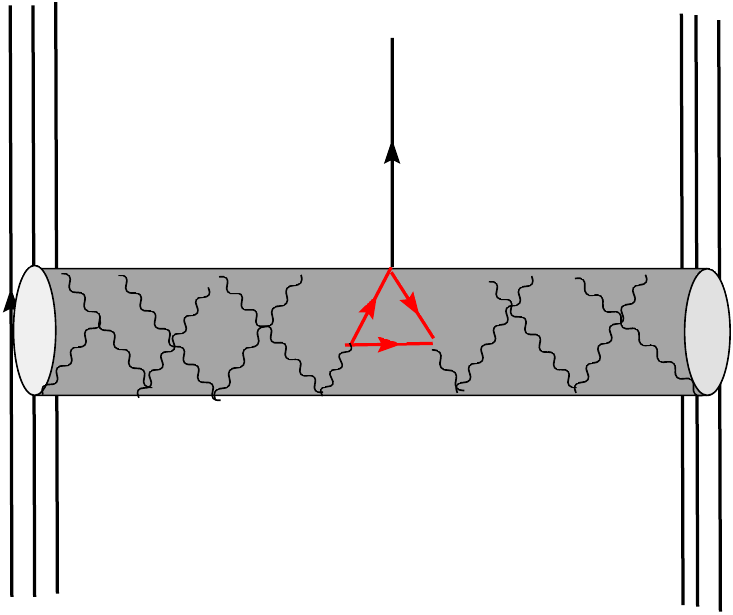}

 \caption{(a) Kinematics for single-Regge limit for  2-to-2 amplitudes, (b) Double-Regge kinematics for  2-to-3 amplitudes.  (c) Cylinder Diagram for large $N_c$ Higgs Production.}
 \label{fig:cylindarHiggs}
\end{figure}

A holographic treatment of Higgs production  amounts to a generalization of our previous $AdS$ treatment for 2-to-2 amplitudes to one for  2-to-3 amplitudes, e.g., from Fig. \ref{fig:cylindarHiggs}a to Fig. \ref{fig:cylindarHiggs}b. A more refined analysis for Higgs production requires a careful treatment for that depicted in Fig. \ref{fig:cylindarHiggs}c.  A particularly useful paper for the diffractive Higgs analysis
is the prior work by Herzog, Paik, Strassler and
Thompson~\cite{Herzog:2008mu} on holographic double diffractive
scattering. In this analysis, one generalizes (\ref{eq:adsPomeronScheme}) to 2-to-3 amplitude where
\begin{equation}
A(s,s_1, s_2, t_1, t_2)  = \Phi_{13}*  \widetilde{\cal K}_P*V_H*  \widetilde {\cal K}_P* \Phi_{24} \; ,
\label{eq:adsDoublePomeronScheme}
\end{equation}
schematically represented by Fig. \ref{fig:cylindarHiggs}b. 
However, a new aspect, not addressed in \cite{Herzog:2008mu},  is the issue of scale invariance breaking.   A proper accounting for a non-vanishing gluon condensate $\langle F^2\rangle$ turns out to be a crucial ingredient in understanding the strength of diffractive Higgs production. We now must pause to realize that in any conformal theory the is no dimensional 
parameter to allow for such a  dimensionful two-graviton-dilaton coupling, $M^2 \phi h_{\mu\nu} h^{\mu \nu}$,  emerging  in an  expansion of the AdS gravity action  if scale invariance is maintained.  However since  QCD is not a conformal theory 
this is just one of many reasons to introduce conformal symmetry breaking.  To model an effective QCD background we will for the most
part introduce two modifications of the pure AdS background: (1) an IR hardwall cut-off 
beyond $z = 1/\Lambda_{qcd}$ to give confinement and  linear static quark potential at large distances and (2) 
a slow deformation in the UV ($z \rightarrow 0$) to model the logarithmic running 
for asymptotic freedom.  Both break conformal invariance, which
as we will argue is required to couple the two gravitons to the dilaton and produce a
Higgs in the central rapidity region.   

\paragraph{Pomeron-Pomeron Fusion Vertex}

We are now in a position to focus on the  Higgs vertex, $V_H$. It is important to stress that our general discussion  in moving from
single-Pomeron exchange processes, (\ref{eq:adsPomeronScheme}), to double-Pomeron exchange, (\ref{eq:adsDoublePomeronScheme}), applies equally well for both
diffractive glueball production and for Higgs production. The
difference lies in how to treat the new central vertex.  For the
production of a glueball, the vertex will be proportional to a
normalizable $AdS$ wave-function.  There will also be an overall factor
controlling the strength of coupling to the external states, e.g., the
Pomeron-Pomeron-glueball couplings.  For Higgs production, on the
other hand, the central vertex, $V_H$, involves a non-normalizable
bulk-to-boundary propagator, appropriate for a scalar external
current. This in turns leads to coupling to a Higgs scalar. This is analogous to the use of a non-normalizable current for $P_{13}(z)$ in equation (\ref{eq:p13}). 

A Higgs scalar in the standard model couples exclusively to the quarks
via Yukawa coupling, which for simplicity we will assume is dominated
by the top quark, with
$   
{\cal L} = - \frac{g}{2 M_W} m_t \; \bar t(x) t(x) \phi_H(x).
$
 Taking advantage of the scale separations between the QCD scale, the Higgs mass and the top quark mass,
$
\Lambda_{qcd} \ll m_H \ll 2 m_t
$,
heavy quark decoupling allows one to replace the Yukawa coupling by direct coupling of Higgs to gluons, which is treated  as an external source in the AdS dictionary.
Consequently $V_H$, in a coordinate representation,  is replaced by the vertex  for two AdS Pomerons fusing
at $(x'_{1\perp},z'_1)$ and $(x'_{2\perp},z'_2)$ and propagating this
disturbance to the $\bar t(x) t(x)$ scalar current at the boundary of
AdS.  The double diffractive Higgs vertex $V_H$ can then be obtained in a two-step process.

First, since the
Yukawa Higgs quark coupling is proportional to the quark mass, it is
dominated by the top quark. Assuming  $m_H \ll m_t$,  this can be
replaced by an effective interaction
by evaluating the two gluon Higgs triangle graph in leading order $O(M_H/m_t)$.
Second, using the AdS/CFT dictionary, the external
  source for $F^a_{\mu \nu}F^a_{\mu \nu}(x)$ is placed at the
  AdS boundary ($z_0 \rightarrow 0$) connecting to the Pomeron fusion vertex
in the interior of $AdS_3$ at ${\bf b}_H=(x'_H,z'_H)$, by a scalar bulk-to-boundary propagator,
$K(x'_H - x_H,z'_H,z_0)$.

We are finally in the position to put all the pieces together. Although we eventually want to go to a coordinate representation in order to perform eikonal unitarization, certain simplification can be achieved more easily in working with the  momentum representation. The Higgs production amplitude, schematically given by (\ref{eq:adsDoublePomeronScheme}), can then be written explicitly as
 \begin{eqnarray*}
A(s,s_1,s_2, t_1,t_2)&\simeq &  \int dz_1 dz  dz_2\; \sqrt{-g_1}\sqrt{-g} \sqrt{-g_2}\;\Phi_{13} (z_1) \\
&\times&   \widetilde {\cal K}_P(s_1,t_1,z_1,z)  \; V_H(q^2 , z)\; \widetilde {\cal K}_P(s_2,t_2,z,z_2)\; \Phi_{24}(z_2)  \;.
\label{eq:adsDoublePomeronHiggs}
\end{eqnarray*}
where  $q^2= -m_H^2$.   For this production vertex,  we will keep  it simple by  expressing it as 
\begin{equation}
 V_H( q^2,z)= V_{PP\phi} K(q^2,z) L_H\; .
 \label{eq:HiggsVertex}
\end{equation}
where  $K(q^2,z)$ is the conventionally normalized bulk to boundary propagator, $V_{PP\phi}$ serves as  an overall coupling from two-Pomeron to $F^2$, and $L$ is  the  conversion factor from $F^2$ to Higgs, i.e.,
$
L_H = L(-m_H^2)\simeq  \frac{\alpha_s g}{24 \pi M_W}.
$
By   treating the central vertex $  V_{PP\phi}$ as a constant, which follows from the super-gravity limit, we have ignored possible additional dependence on  $\kappa$, as well as that on $t_1$ and $t_2$. This approximation gives an explicit
factorizable form for Higgs production. 

\paragraph{Strategy for Phenomenological Estimates}

As a first step in making a phenomenological estimate for the cross section, we ask how the central vertex, $V_H$, or equivalently,  $ V_{PP\phi}$,  via (\ref{eq:HiggsVertex}),  can be normalized, following  the approach of Kharzeev and Levin~\cite{Kharzeev:2000jwa} based on the analysis of trace anomaly. 
 We also show how
one can in principle use the elastic scattering to normalize the bare BPST Pomeron coupling to external protons and the 't Hooft 
coupling $g^2N_c$.

We start from Eq. (\ref{eq:adsDoublePomeronHiggs}). When nearing the respective tensor poles at $t_1\simeq  m_0^2$ and $ t_2\simeq  m_0^2$,  the amplitude can be expressed as
\begin{eqnarray}
A(s,s_1,s_2, t_1,t_2)&\simeq&g_{13}\; \frac{  \Gamma_{GGH} \;  s^2  }{(t_1-m_0^2)(t_2-m_0^2)  }   \; g_{24}
   \label{eq:2to3KL}
\end{eqnarray}
We have performed the $z_1$ and $z_2$ integrations,  and   have also made use of the fact that $s_1s_2\simeq \kappa\; s\simeq m_H^2 s $. 
Here $\Gamma_{GGH}$ is the effective on-shell glueball-glueball-Higgs coupling, which  can also be expressed as
$
 \Gamma_{GGH} =L_H F ( - m^2_H)
$
where $L_H=    \frac{\alpha_s g}{24 \pi M_W}$ and     $ F $  is a scalar form factor
$
F(q^2) = <G, ++,q_1| F^a_{\mu\nu} F^a_{\mu\nu}(0) | G,--,q_2>.
$ 
That is, in the high energy Regge limit, the dominant contribution comes from  
 the maximum helicity glueball state~\cite{Brower:2006ea}, with $\lambda = 2$.  In this limit,  this form factor,  is given by
the overlap of the dilaton bulk to boundary propagator
\bea
F(q^2) &=&(\alpha' m_H^2)^2   V_{PP\phi} \int dz \sqrt{-g(z)} e^{-4A(z)} \phi_G(z) K(q,z) \phi_G(z)
\label{eq:formfactor}
\eea
  What remains to be specified is the overall normalization, $F(0)$.

We next follow  D. Kharzeev and E. M. Levin  \cite{Kharzeev:2000jwa}, who noted  that, from the SYM side,  $F(q^2)$ at $q^2=0$, can be considered as the glueball condensate. Consider matrix elements of  the trace-anomaly  between two states, $|\alpha(p)>$ and $|\alpha'(p')>$, with four-momentum transfer $q=p-p'$. In particular, for a single particle state of a tensor glueball $|G(p)>$, this leads to $<G(p)|  {\Theta}^\alpha_\alpha|G(p')> = \frac{\widetilde \beta}{2 g} <G(p)| F^a_{\mu \nu} F^{a \mu \nu} | G(p')  >$. 
At $q = 0$, the forward matrix element of the trace of the energy-momentum tensor is given simply by  the mass of the relevant tensor glueball, with  $<G|{\Theta}^\alpha_\alpha | G  > = M_G^2$, this directly yields
\be
F(0) = <G| F^a_{\mu \nu} F^{a \mu \nu} | G  >  =-  \frac{4\pi  M_G^2}{3 \widetilde \beta }
\label{eq:FF}
\ee
where $\widetilde \beta = - b\alpha_s/(2 \pi)$, $b = 11 - 2n_f/3$, for $N_c=3$.  In what follows, we will use $n_f=3$.   Note that heavy quark contribution is not included in this  limit.  Since the conformal scale breaking is due the running coupling constant in QCD, 
there is apparently a mapping between QCD scale breaking and breaking of
the AdS background  in the IR, which  gives a finite mass to the glueball
and to give a non-zero contribution to the gauge condensate.

Let us turn next to the non-forward limit.  We  accept the fact that, in the physical region where $t<0$ and small, the cross sections typically have an exponential form, with a logarithmic slope which is mildly energy-dependent. We therefore approximate all amplitudes in the near forward region where $t<0$ and small,
$
A(s,t) \simeq  e^{B_{eff}(s)\; t/2}\;  A(s,0)
$ 
where $B_{eff}(s)$ is a smoothly slowly increasing function of s, (we expect it to be logarithmic).
We also assume, for $t_1<0$, $t_2<0$ and small, the Higgs production amplitude is also strongly damped so that
\be
A(s,s_1,s_2, t_1, t_2)  \simeq    e^{B'_eff(s_1) \; t_1/2}  e^{B'_eff(s_2)\; t_2 /2   }  \; A(s,s_1,s_2, t_1\simeq 0, t_2\simeq 0)
\label{eq:higgs}
\ee
We also assume $B'_{eff}(s)\simeq B_{eff}(s) + {\rm  b}$.   With these, both the elastic, the total pp cross sections and the Higgs production cross section can now be evaluated.  Various cross sections will of course depend on the unknown slope parameter, $B_{eff}$, which can at best be estimated based  on prior experience with diffractive estimates. One can relate $B_{eff}$ directly in terms of the experimentally smooth dimensionless ratio,
$
R_{el}(s) = {\sigma_{el}}/{\sigma_{total}}  = \frac{ (1+\rho^2) \sigma_{total}(s) }{16\pi B_{eff} (s) }.
$
Upon squaring the amplitude, $A(s,s_1,s_2, t_1, t_2)$, (\ref{eq:higgs}), the double-differential cross section for Higgs production can now be obtained. 
 After integrating over $t_1$ and $t_2$ and using the fact that, for $m_H^2$ large  $s \simeq  {s_1 s_2}/{m_H^2}$,
one finds
\begin{eqnarray}
\frac{d\sigma}{dy_H } &\simeq &(1/\pi) \times C'   \times |   \Gamma_{GGH}(0)/\widetilde m^2|^2  \times \frac{\sigma( s)}{\sigma(m_H^2)} \times R^2_{el}(m_H \sqrt s )
\end{eqnarray}
The value of the above result is model dependent, and with our model is $\sim 1 pb$. This is of the same order as estimated in \cite{Kharzeev:2000jwa}.  However, as also pointed in \cite{Kharzeev:2000jwa}, this should be considered as an over-estimate. The major source of suppression will come from absorptive correction, which can lead to a central production cross section in the  femtobarn  range. A lot of details have been glossed over in the above derivation, see~\cite{Brower:2012mk}.

\paragraph{Conclusions:}

We have presented the phenomenological application of the AdS/CFT correspondence to the study of high energy diffractive scattering
for QCD.  Fits to the HERA DVCS data at small x demonstrate that the strong coupling BPST Graviton/Pomeron~\cite{Brower:2006ea}  does allow for a
very good description of diffractive DVCS with few phenomenological parameters, the principal one being the intercept to the bare Pomeron fit to be $j_0 \simeq   1.2-1.3$.
Encouraged by this, we plan to undertake a fuller study of several closely related diffractive process: total and elastic cross sections, DIS, virtual photon production, vector meson production
and double diffractive production of heavy quarks. The goal is that by over-constraining the basic AdS building blocks of diffractive
scattering, this framework will give 
a compelling phenomenological prediction for the double diffractive production of the Higgs in the standard model to aid in the analysis of LHC data.

\paragraph*{Acknowledgments}

The work of M.S.C. and M.D. was partially funded by grants PTDC/FIS/ 099293/2008 and CERN/FP/ 116358/2010.
\emph{Centro de F\'{i}sica do Porto} is partially funded by FCT. The work of M.D. is supported by the FCT/Marie Curie Welcome II program.  The
work of R.C.B.  was supported by the Department of Energy under
contract~DE-FG02-91ER40676, and that of  C.-IT.  by the
Department of Energy under contract~DE-FG02-91ER40688, Task-A. R.B. and C.-IT. would like to thank the Aspen Center
for Physics for its hospitality during the early phase  of this work.

\paragraph*{References}


\begin{thebibliography}{99}

\bibitem{Brower:2006ea}
  R.~C.~Brower, J.~Polchinski, M.~J.~Strassler, C.~-I~Tan,
  JHEP {\bf 0712}, 005 (2007).

\bibitem{Costa:2012fw}
  M.~S.~Costa and M.~Djuric,
  Phys.\ Rev.\ D {\bf 86} (2012) 016009
  [arXiv:1201.1307 [hep-th]].


\bibitem{Brower:2012mk}
  R.~C.~Brower, M.~Djuric and C.~-I~Tan,
  JHEP {\bf 1209} (2012) 097
  [arXiv:1202.4953 [hep-ph]].

\bibitem{Brower:2010wf}
  R.~C.~Brower, M.~Djuri\'c, I.~Sarcevic, C.~-I~Tan,
  JHEP {\bf 1011}, 051 (2010).

\bibitem{Brower:2011dx}
  R.~C.~Brower, M.~Djuric, I.~Sarcevic and C.~-ITan,
  arXiv:1106.5681 [hep-ph].

\bibitem{Saturation}L.~Cornalba and M.~S.~Costa, 
Phys.\ Rev.\  D {\bf 78} (2008) 096010,
\texttt{arXiv:0804.1562 [hep-ph]}.

\bibitem{Maldacena:1997re}
  J.~M.~Maldacena,
  Adv.\ Theor.\ Math.\ Phys.\  {\bf 2}, 231 (1998)

\bibitem{Witten:1998qj}
  E.~Witten,
  Adv.\ Theor.\ Math.\ Phys.\  {\bf 2}, 253-291 (1998).




\bibitem{low}F.E. Low, \Journal{\PRD}{12}{163}{1975}

\bibitem{nussinov}S. Nussinov, \Journal{\PRL}{34}{1286}{1975}

\bibitem{Cornalba:2010vk}
  L.~Cornalba, M.~S.~Costa, J.~Penedones,
  Phys.\ Rev.\ Lett.\  {\bf 105}, 072003 (2010).





\bibitem{Gao:2009se}
  J.~H.~Gao and B.~W.~Xiao,
Phys.\ Rev.\ D {\bf 81} (2010) 035008,
 \texttt{arXiv:0912.4333 [hep-ph]}.
  
\bibitem{Marquet:2010sf}
  C.~Marquet, C.~Roiesnel, S.~Wallon,
  JHEP {\bf 1004 } (2010)  051,
 \texttt{arXiv:1002.0566 [hep-ph]}.
    
\bibitem{Nishio:2011xz}
  R.~Nishio and T.~Watari,
  Phys.\ Rev.\ D {\bf 84} (2011) 075025,
 \texttt{arXiv:1105.2999 [hep-ph]}.

  R.~Nishio and T.~Watari, 
   \texttt{arXiv:1105.2907 [hep-ph]}.



\bibitem{Polchinski:2002jw}
  J.~Polchinski, M.~J.~Strassler,
  JHEP {\bf 0305}, 012 (2003).

\bibitem{Brower:2007qh}
  R.~C.~Brower, M.~J.~Strassler, C.~-I~Tan,
  JHEP {\bf 0903}, 050 (2009).

\bibitem{Brower:2007xg}
  R.~C.~Brower, M.~J.~Strassler, C.~-I~Tan,
  JHEP {\bf 0903}, 092 (2009).

\bibitem{Brower:2008cy}
  R.~C.~Brower, M.~Djuric and C-I~Tan,
  JHEP {\bf 0907}, 063 (2009).

\bibitem{Cornalba:2006xm}
  L.~Cornalba, M.~S.~Costa, J.~Penedones, R.~Schiappa,
  Nucl.\ Phys.\  {\bf B767}, 327-351 (2007).

\bibitem{Cornalba:2008qf}
  L.~Cornalba, M.~S.~Costa, J.~Penedones,
  JHEP {\bf 0806}, 048 (2008).


\bibitem{Cornalba:2007zb}
  L.~Cornalba, M.~S.~Costa, J.~Penedones,
  JHEP {\bf 0709}, 037 (2007).

\bibitem{Chekanov:2008vy}
  S.~Chekanov {\it et al.}  [ZEUS Collaboration],
  JHEP {\bf 0905} (2009) 108,
\texttt{arXiv:0812.2517 [hep-ex]}.

\bibitem{:2009vda}
  F.~D.~Aaron {\it et al.}  [H1 Collaboration],
  Phys.\ Lett.\  B {\bf 681} (2009) 391,
\texttt{arXiv:0907.5289 [hep-ex]}.

  F.~D.~Aaron {\it et al.}  [H1 Collaboration],
  Phys.\ Lett.\ B {\bf 659} (2008) 796,
\texttt{arXiv:0709.4114 [hep-ex]}.

\bibitem{Kharzeev:2000jwa}
  D.~Kharzeev and E.~Levin,
  Phys.\ Rev.\  D {\bf 63}, 073004 (2001)
  [arXiv:hep-ph/0005311].

\bibitem{Herzog:2008mu}
  C.~P.~Herzog, S.~Paik, M.~J.~Strassler and E.~G.~Thompson,
  JHEP {\bf 0808}, 010 (2008)
  [arXiv:0806.0181 [hep-th]].



\end{thebibliography}
\end{document}